\documentclass[onecolumn,floatfix,aps,nofootinbib,showpacs, showkeys]{revtex4-2}
\usepackage{bbold}
\usepackage[dvips]{epsfig}
\usepackage[english]{babel}
\usepackage{bbm}
\usepackage{verbatim}
\usepackage{array}
\usepackage{amsmath}
\usepackage{multirow}
\usepackage{amsfonts}
\usepackage{amssymb}
\usepackage{leading}
\usepackage[dvipsnames]{xcolor}
\usepackage{subeqnarray}
\usepackage{setspace}
\usepackage{placeins}
\usepackage{float}
\usepackage{indentfirst} 

\definecolor{myblue}{rgb}{0,0,0.8}
\usepackage[colorlinks=true
,urlcolor=myblue	
,anchorcolor=myblue
,citecolor=myblue
,filecolor=myblue
,linkcolor=myblue
,menucolor=myblue
,linktocpage=true 
]{hyperref}

\usepackage{gensymb}

\def\p{\mbox{\boldmath$\displaystyle\boldsymbol{p}$}}
\newcommand{\gdualn}[1]{\overset{\:{}^{{}^{\boldsymbol{\neg}}}}{\smash[t]{#1}}}
\def\C{\mbox{\boldmath$\displaystyle\mathbb{C}$}}
\def\R{\mbox{\boldmath$\displaystyle\mathbb{R}$}}
\def\s{\mbox{\boldmath$\displaystyle\boldsymbol{\sigma}$}}
\def\gb{\mbox{\boldmath$\displaystyle\boldsymbol{\gamma}$}}
\def\0{\mathbb{0}}
\def\a{\mbox{\boldmath$\displaystyle\boldsymbol{a}$}}

\begin{document}

\title{Propagators Beyond The Standard Model}
\author{Rodolfo Jos\'e Bueno Rogerio$^{1}$}\email{rodolforogerio@gmail.com}
\author{Luca Fabbri$^{2}$}\email{fabbri@dime.unige.it}
\affiliation{$^{1}$Institute of Physics and Chemistry, Federal University of Itajub\'a , Itajub\'a, Minas Gerais, 37500-903, BRAZIL\\
$^{2}$DIME, Sez. Metodi e Modelli Matematici, Universit\`{a} di
Genova, via all'Opera Pia 15, 16145 Genova, ITALY}


\begin{abstract}
\noindent{\textbf{Abstract.}}
In this paper, we explore the field propagator with a structure that is general enough to encompas both the case of newly-defined mass-dimension $1$ fermions and spin-$1/2$ bosons. The method we employ is to define a map between spinors of different Lounesto classes, and then write the propagator in terms of the corresponding dual structures. 
\end{abstract}
\maketitle
\section{Introduction}\label{intro}
Following to Lounesto's classification \cite{lounestolivro}, spinor fields can be split into six major classes according to which spinor bilinear vanishes: the first three classes are the so-called \emph{regular} spinors and they contain those spinor fields whose bilinear scalar and pseudo-scalar are generally not identically zero; the last three clsses are the so-called \emph{singular} spinors and they contain those spinor fields whose bilinear scalar and pseudo-scalar are both identically null. For our purposes it will be immaterial to proceed into the sub-classification of the first three classes: it will suffice to say that they contain the Dirac spinor fields. The three classes of singular spinors, also called \emph{flag-dipole} spinors, are sub-classified as follows: the first class contains the singular spinors that are not subject to further restrictions; the second class contains the singular spinors whose bilinear axial-vector vanishes; the third class contains the singular spinors whose bilinear tensor momentum vanishes. Respectively, these last three classes contain the \emph{flag-dipole} spinors in strict sense, the \emph{flagpole} spinors, exhausted by the Majorana spinor, and the \emph{dipole} spinors, exhausted by the Weyl spinor. Schematically, we have:
\begin{itemize}
\item not both $\overline{\psi}\psi$ and $\overline{\psi}\gamma^{5}\psi$ are zero: regular, Dirac spinor fields (classes I, II, III);
\item $\overline{\psi}\psi\!=\!0$ and $\overline{\psi}\gamma^{5}\psi\!=\!0$: singular, flag-dipoles (classes IV, V, VI):
  \begin{itemize}
   \item no further restriction: flag-dipoles in strict sense (class IV),
   \item $\overline{\psi}\gamma^{i}\gamma^{5}\psi\!\equiv\!0$: flagpoles, Majorana spinors (class V),
   \item $\overline{\psi}\sigma^{ij}\psi\!\equiv\!0$: dipoles, Weyl spinors (class VI).
  \end{itemize}
\end{itemize}
The Dirac spinor fields are the basic building-blocks of the standard model of particles physics. Flag-dipoles, albeit well established mathematically, have never been used in physical theories, at least so far, but the two sub-classes of Weyl and Majorana spinors are also basic constituents of many extensions of the standard model of particle physics, encompassing the problem of neutrino masses, dark matter \cite{mdobook}, as well as superstring theory, just to name the most famous members of the ``Beyond the Standard Model'' (BSM) sector.

Quite recently, however, the BSM sector saw the inclusion of new theoretical elements. These new elements are spinor fields still belonging to the $(1/2, 0)\oplus(0, 1/2)$ representation but having mass-dimension $1$, that is spinor fields whose quantum field propagator does not have the structure of a fermion propagator (mass-simension $3/2$) but that of a boson propagator (mass-dimension $1$) \cite{dharamboson,dharamnewfermions}. Conversely, one could also think to introduce a theory of scalar fields with fermionic propagator (mass-dimension $3/2$). At this point, we call attention to the  fact that spinors are neither fermionic nor bosonic. They can be used to construct quantum fields, and these fields may be fermionic or bosonic. The kind of field is intrinsically connected according the definition of the dual structure, the reader may refer to the well posed definitions in \cite{dharamboson,dharam2022spin}.

In the present work, we aim at establishing the most general form of propagators for these fields, whether they are fermions with bosonic propagators or bosons with fermionic propagators.

For this, the method we employ is build on defining a map between spinors, and then write the propagator in terms of the physical information that connects spinors of the different classes as given in \cite{rodolfobosonsphases}. Our program, hence, splits into two fundamental steps: define how the spinors relate to each other, and then, see how the dual structures are connected and how this affects the spin-sum relations.

We will show that propagators can be written as a momentum-dependent term plus another term proportional to the identity, as one would expect in quantum field theoretical constructions.

The paper is organized as follows: in the first section we explore the  mathematical aspects of the mass-dimension $1$ fermions and the spin-$1/2$ bosonic spinors. In Sect. \ref{generalpropagatorboson} we explore a mathematical device that allows us to write the propagator in a general form, employing a mapping procedure among spinors to define the amplitude of propagation in all cases, and evincing a possible connection among propagators given in terms of the chiral symmetry. Finally, in sect.\ref{concluding}, we conclude with some general remarks.     
\section{Interpreting the propagators}
We begin by introducing a quantum field in its most general form, with $\lambda_i(\p)$ (for particles) and $\chi_i(\p)$ (representing anti-particle) as its expansion coefficients: it is given by
\begin{equation}
\mathfrak{f}(x) \stackrel{\mathrm{def}}{=}
\int\frac{d^3 p}{(2\pi)^3}
\frac{1}{\sqrt{2 \upsilon E(\p)}}
\bigg[
\sum_{i=particles} {c}_i(\p)\lambda_i(\p) e^{-i p\cdot x}+ \sum_{i=anti-particles} d^\dagger_i(\p)\chi_i(\p) e^{i p\cdot x}\bigg],\label{eq:fieldb}
\end{equation}
and 
\begin{equation}
\gdualn{\mathfrak{f}}(x) \stackrel{\mathrm{def}}{=}
\int\frac{d^3 p}{(2\pi)^3}
\frac{1}{\sqrt{2 \upsilon E(\p)}}
\bigg[
\sum_{i=particles} c^\dagger_i(\p)\gdualn{\lambda}_i(\p) e^{i p\cdot x}+ \sum_{i=ianti-particles} d_i(\p)\gdualn{\chi}_i(\p) e^{-i p\cdot x}\bigg],
\end{equation}
as its adjoint. The $\upsilon$ is a free parameter related to the type of field we are considering: it is $\upsilon = m$ for fields of bosonic type (second-order differential field equations) and $\upsilon=1$ for fields of fermionic type (first-order differential field equations). Generally speaking, one could justify this on the grounds of a dimensional analisys.

To keep the development general, we shall not fix the statistics, neither to be fermionic 
\begin{equation}
\left\{c_i(\p),c^\dagger_j(\p)\right\} =(2 \pi)^3 \delta^3(\p-\p^\prime)\delta_{ij}, \quad \left\{c_i(\p), c_j(\p^\prime)\right\} = 0 =
 \left\{c^\dagger_i(\p), c^\dagger_j(\p^\prime)\right\}, \label{anticomutador}
\end{equation}
and nor bosonic
\begin{equation}
\left[c_i(\p),c^\dagger_j(\p)\right] =(2 \pi)^3 \delta^3(\p-\p^\prime)\delta_{ij}, \quad \left[c_i(\p), c_j(\p^\prime)\right] = 0 =
 \left[c^\dagger_i(\p), c^\dagger_j(\p^\prime)\right] \label{comutador}
\end{equation}
(similar relations are also assumed for the remaining operators $d_i(\p)$ and $d_i^\dagger(\p)$).

Looking for a way to determine the (fermionic or bosonic) statistics for the quantum fields $\mathfrak{f}(x)$ and $\gdualn{\mathfrak{f}}(x)$, we must consider two events, $x$ and $x'$, and then note that the amplitude of propagation from $x$ to $x'$ is given by the following relation
\begin{align}
\mathcal{A}(x\to x^\prime)   =  \xi \Big(\underbrace{\langle\hspace{3pt}\vert
\mathfrak{f}(x^\prime)\gdualn{\mathfrak{f}}(x)\vert\hspace{3pt}\rangle \theta(t^\prime-t)
\pm  \langle\hspace{3pt}\vert
\gdualn{\mathfrak{f}}(x) \mathfrak{f}(x^\prime)\vert\hspace{3pt}\rangle \theta(t-t^\prime)}_{\langle\hspace{4pt}\vert \mathfrak{T} (\mathfrak{f}(x^\prime) \gdualn{\mathfrak{f}}(x))\vert\hspace{4pt}\rangle}\Big)\label{amplitude1}
\end{align}
in which the plus sign holds for bosons and the minus sign for fermions. The arbitrary constant $\xi\in\C$ is to be further determined by a normalisation condition. $\mathfrak{T}$ is the well known time-ordering operator.

As for the two vacuum-expectation-values that appear in $\mathcal{A}(x\to x^\prime) $ we have
 \begin{align}
\langle\hspace{3pt}\vert
\mathfrak{f}(x^\prime)\gdualn{\mathfrak{f}}(x)\vert\hspace{3pt}\rangle  & =\int\frac{d^3p}{(2 \pi)^3}\left(\frac{1}{2 \upsilon E(\p)}\right)
 e^{-ip\cdot(x^\prime-x)}
 \sum_{particle}\lambda^{f/b}_i(\p)
\gdualn\lambda^{f/b}_i(\p) \label{amplitudeP-S}
\\
\langle\hspace{3pt}\vert
\gdualn{\mathfrak{f}}(x) \mathfrak{f}(x^\prime)\vert\hspace{3pt}\rangle 
  & =  \int\frac{d^3p}{(2 \pi)^3}\left(\frac{1}{2 \upsilon E(\p)}\right)
 e^{ip\cdot(x^\prime-x)}
 \sum_{anti-particle}\!\!\!\!\!\!\chi^{f/b}_i(\p)
\gdualn\chi^{f/b}_i(\p) .\label{amplitudeP-A}
\end{align}

Thus, we can re-write the spin-sum relations in the right-hand side of Eqs.\eqref{amplitudeP-S} and \eqref{amplitudeP-A}. Summing over all helicities, for regular fermionic particles \cite{rjfermionicfield} we get
\begin{eqnarray}
\sum_{particle}\lambda^{f}(\p)\gdualn\lambda\;^{f}(\p) &=& \gamma_{\mu}p^{\mu}+ m\mathbbm{I}_{(\alpha,\beta)}, \label{spinsums1}
\\
\sum_{anti-particle}\!\!\!\!\!\!\chi^{f}(\p)\gdualn\chi\;^{f}(\p) &=& \gamma_{\mu}p^{\mu}- m\mathbbm{I}_{(\alpha,\beta)},\label{spinsums2}
\end{eqnarray}
in which $\mathbbm{I}_{(\alpha,\beta)} =\mathrm{diag}(e^{i(\alpha-\beta)}, e^{i(\alpha-\beta)}, e^{-i(\alpha-\beta)}, e^{-i(\alpha-\beta)})$ is a diagonal matrix written in terms of the spinor phase factors. For the Dirac case, yielding the parity relation $\mathcal{P}\psi = \pm \psi$, we have $\mathbbm{I}_{(\alpha,\beta)} \to \mathbbm{I}$. For flag-dipoles (including in particular dipoles, flagpoles, and thus also mass-dimension $1$ fermions \cite{mdobook,dharamnewfermions,chengtipo42021}), we get 
\begin{eqnarray}
\sum_{particle}\lambda^{f}(\p)\gdualn\lambda\;^{f}(\p) &=& 2m\mathbbm{I},
\\
\sum_{anti-particle}\!\!\!\!\!\!\chi^{f}(\p)\gdualn\chi\;^{f}(\p) &=& -2m\mathbbm{I}.
\end{eqnarray}

The spin-half bosonic counterpart yields instead \cite{dharamboson}
\begin{eqnarray}
\sum_{particle}\lambda^{b}_{i}(\p)\gdualn\lambda\;^{b}_{i}(\p) &=& a_{\mu}p^{\mu}+ m\mathbbm{I}, \label{ssbosonpart}
\\
\sum_{anti-particle}\!\!\!\!\!\!\chi^{b}_{i}(\p)\gdualn\chi\;^{b}_{i}(\p) &=& -a_{\mu}p^{\mu}+ m\mathbbm{I}. \label{ssbosonanti}
\end{eqnarray}
We highlight here the fundamental effects of the spinorial dual on the above structures. For Dirac spinors and mass dimension one fermions one has a complete set of orthonormal relations, with $\pm 2m$, the upper sign for particles and the lower sign for anti-particles. For the spin-half bosons, the orthonormal relation is always positive, and this fact brings consequences for the propagator computation, as we have shown in the appendix.

The relationship among $\gamma_{\mu}$ and $a_{\mu}$ matrices reads \cite{dharamboson} 
\begin{eqnarray}\label{gamma-a}
\gamma_{\mu}=ia_5 a_{\mu}, \quad a_{\mu} = i\gamma_5\gamma_{\mu},
\end{eqnarray} 
where, in chiral representation, we have
\begin{eqnarray}
a_0 = i \left(\begin{array}{cc}
0 & \mathbbm{I}\\
-\mathbbm{I} & 0
\end{array}\right),\quad
\a  = i \left(\begin{array}{cc}
0 &\s\\
\s & 0
\end{array}\right),
\end{eqnarray}
and
\begin{eqnarray}
\gamma_0 = 
\left(\begin{array}{cc}
0 & \mathbbm{I}\\
\mathbbm{I} & 0
\end{array}\right),\quad
\gb  =  \left(\begin{array}{cc}
0 &\s\\
- \s & 0
\end{array}\right),
\end{eqnarray}
being $\s$ the Pauli matrices. Since both $\gamma_{\mu}$ and $a_{\mu}$ satisfy the same algebra, there exists a similarity transformation that can bring $a_{\mu}$ to $\gamma_{\mu}$ and vice versa. Nonetheless, such a transformation takes one out of the Weyl representation, in view of this facts $a_{\mu}$ and $\gamma_{\mu}$  are physically distinct.

The isomorphism presented in \eqref{gamma-a} was first noticed in \cite[pages 447-448]{brauer} and later scrutinized in \cite[Chapter 17]{bosonalgebrabook}.

The two Heaviside step functions of equation (\ref{amplitude1}) can now be replaced by their integral representations
\begin{align}
\theta(t^\prime-t) &=  -\frac{1}{2\pi i}\int\text{d}\omega
\frac{e^{i \omega (t^\prime-t)}}{\omega- i \epsilon}, \\
\theta(t-t^\prime) &=  -\frac{1}{2\pi i}\int\text{d}\omega
\frac{e^{i \omega (t-t^\prime)}}{\omega- i \epsilon},
\end{align}
where $\epsilon, \omega \in\R$. Thus, with such quantities, the amplitude of propagation reads 
\begin{eqnarray}\label{propagador666}
\mathcal{A}_{x\rightarrow x^\prime}
= -\xi \mathop{\mathrm{lim}}\limits_{\epsilon\rightarrow 0^+}
 \int\frac{\mathrm{d}^3p}{(2\pi)^3}\frac{1}{2E(\boldsymbol{p})\upsilon}
\int\frac{\mathrm{d}\omega}{2\pi i}&\times &
\bigg[\frac{\sum_{particle}\lambda^{f/b}(\p)\gdualn\lambda\;^{f/b}(\p)}{\omega - i\epsilon}\mathrm{e}^{i(\omega-E(\boldsymbol{p})(t^\prime -t)}\,\mathrm{e}^{i 
\boldsymbol{p}.(\mathbf{x}^\prime-\mathbf{x})}\nonumber\\
&\pm&
\frac{\sum_{anti-particle}\chi^{f/b}(\p)\gdualn\chi\;^{f/b}(\p)}{\omega - i\epsilon}\mathrm{e}^{-i(\omega-E(\boldsymbol{p})(t^\prime -t)}\,\mathrm{e}^{i
\boldsymbol{p}.(\mathbf{x}^\prime-\mathbf{x})}\bigg].
\end{eqnarray}

Shifting $\omega \to p_0 = -\omega+E(\p)$ in the first integral and $\omega \to p_0 = \omega- E(\p)$ in the second integral, and taking into account the definition of the spin-sum relations above, we can have \eqref{propagador666} shifted into
\begin{eqnarray}\label{general-propagator-form}
\mathcal{A}(x\to x^\prime)
= i\xi \mathop{\mathrm{lim}}
\limits_{\epsilon\rightarrow 0^+}
\int \frac{\mathrm{d}^4 p}{(2\pi)^4} \frac{1}{2 E(\boldsymbol{p})\upsilon} 
\mathrm{e}^{-i p_\mu(x^{\prime\mu} - x^\mu)}
\bigg[ \frac{ \sum_{particle}\lambda^{f/b}(\p)\gdualn\lambda\;^{f/b}(\p)}{E(\p) - p_0 - i\epsilon}
\pm
\frac{\sum_{anti-particle}\chi^{f/b}(-\p)\gdualn\chi\;^{f/b}(-\p)} {E(\p) + p_0 - i\epsilon}\bigg].\nonumber\\
\end{eqnarray}
where we leave the two factors $\upsilon$ and $\xi$ to be fixed by the type of spinor and its dynamical character (that is, again, by the differential order of the field equations). 

In the next sections, we will derive the amplitude of propagation and also investigate the structure of the propagators for the quite recent theories that are in the scope of BSM theories.
\section{The most general structure}\label{generalpropagatorboson}
Bearing in mind the existence, besides the one presented in \cite{dharamboson}, of other classes of spinors endowed with bosonic traces, that is the previously-mentioned bosons of spin-$1/2$ \cite{rodolfobosonsphases}, we start assuming the possibility of mapping the spin-half bosons presented in \cite{rodolfobosonsphases} and also the possibility to define a mapping program for mass-dimension $1$ fermions. The protocol developed here points toward mapping fermions into fermions and bosons into bosons. Any ``\emph{mixed}'' mapping procedure, taking fermions into bosons, is more involved and will be treated in a following work.

Let $\lambda$ and $\psi$ be spinor fields (either fermionic or bosonic). Thus, one may define the following mapping relation
\begin{equation}\label{mapusual1}
\lambda = \mathbf{M}\psi
\end{equation} 
where $\mathbf{M}$ is a $4\times 4$ diagonal matrix, denoted by $\mathbf{M}=\mathrm{diag}(\alpha,\alpha,\beta ,\beta )$ where $\alpha, \beta\in\mathbb{C}$ \cite{rodolfobosonsphases}. This most general form of the $\mathbf{M}$ operator opens up the possibility to access all classes, mapping spinors according to the choice of the parameter $\alpha$ and $\beta$, as it can be seen in \cite{rodolfoconstraints,rodolfobosonsphases}. To ensure an invertible map, we expect that $\mathbf{M}^{-1}$ exists, yielding 
\begin{equation}\label{mapusual2}
\psi = \mathbf{M}^{-1}\lambda.
\end{equation}

Now suppose $\lambda$ and $\psi$ carry a generalized dual structure leading to the following dual definition
\begin{equation}\label{map2}
\stackrel{\neg}{\lambda} = [\Xi_{\lambda}\lambda]^{\dag}\eta_0,  
\end{equation}
and
\begin{equation}\label{map2.2}
\stackrel{\neg}{\psi} = [\Xi_{\psi}\psi]^{\dag}\eta_0, 
\end{equation}
where $\Xi$ stands for a $4\times 4$ matrix with $\Xi^2=\mathbbm{I}$. Then, $\Xi^{-1}$ indeed exists and $\eta_{\mu}=\gamma_{\mu}$ for fermions while $\eta_{\mu}=a_{\mu}$ for the spin-$1/2$ bosons. The dual structures is generalized compared to the Dirac dual in the sense that it gives the possibility to define spinor bilinear that can be non-zero even for the spinors whose bilinears were zero when computed in the Dirac case. So for instance, a spinor belongin to a class V Majorana type must necessariky have $\overline{\psi}\psi=0$ when the dual is a Dirac dual, but it can have $\stackrel{\neg}{\lambda}\lambda\neq0$ for another type of dual structure. We point the interested readers to \cite{spinorialduals2022} for a complete review on spinorial dual theory.

From \eqref{mapusual1}, \eqref{mapusual2}, \eqref{map2} and \eqref{map2.2} we are able to define the mapped adjoint structure, furnishing
\begin{equation}
\stackrel{\neg}{\psi} = \stackrel{\neg}{\lambda}\eta_0{(\Xi_{\psi}^{\dag}\mathbf{M}^{\dag}\Xi_{\lambda}^{\dag})}^{-1}\eta_0,
\end{equation}
and 
\begin{equation}
\stackrel{\neg}{\lambda} = \stackrel{\neg}{\psi}\eta_0(\Xi_{\psi}^{\dag}\mathbf{M}^{\dag}\Xi_{\lambda}^{\dag})\eta_0.
\end{equation}

The usual elements of the Clifford algebra basis 
\begin{equation}\label{cliffordbasis}
\Gamma = \lbrace \mathbbm{I}, \eta_{\mu}, \eta_5, \eta_5 \eta_{\mu}, \eta_{\mu}\eta_{\nu}\rbrace,
\end{equation}
are taken to
\begin{equation}\label{cliffordbasisdeformed}
\Gamma \rightarrow \tilde{\Gamma} = \eta_0\Xi_{\lambda}^{\dag}\mathbf{M}^{\dag}\Xi_{\psi}^{\dag}\eta_0\Gamma\mathbf{M}.
\end{equation}

Because the mapping is invertible, all physical information of $\lambda$ may be obtained from $\psi$ and vice-versa. Also note that the physical information is obtained from $\mathbf{M}$ and $\Xi$ operators. We highlight that equation \eqref{cliffordbasisdeformed} is useful to re-write and connect different amplitude of propagation, as it will be clear in the next steps.

The amplitude of propagation in \eqref{general-propagator-form} can be splited as follows 
\begin{equation}
\mathcal{A}(x-x^{\prime}) = \mathcal{A}(x-x^{\prime})_{particle}\pm\mathcal{A}(x-x^{\prime})_{anti-particle}.
\end{equation} 

So far we have defined the most general amplitude of propagation. Employing the spinor mapping, an easy way to compute a specific amplitude of propagation is accomplished by inserting the relations obtained in \eqref{cliffordbasisdeformed} in the spin-sums that appears in the general form of \eqref{general-propagator-form}. So, if we know the propagator of a given theory, we are able to convert it into the propagator of the mapped theory. The essential element is just the definition of the parameters $\mathbf{M}$ and $\Xi$ that do the conversion, and therefore the mapping.

Thus, some simple calculations lead to a general (fermionic or bosonic) mapped amplitude of propagation
\begin{equation}\label{propagadormapeado}
\mathcal{A}_{\mathbf{M},\Xi}(x-x^{\prime}) = \mathbf{M}\mathcal{A}(x-x^{\prime})\eta_0\Xi_{\lambda}^{\dag}\mathbf{M}^{\dag}\Xi_{\psi}^{\dag}\eta_0.
\end{equation}   
Notice that once we have defined the amplitude $\mathcal{A}(x-x^{\prime})$ for any mass-dimension $1$ spinor or spin-$1/2$ boson, the $\mathcal{A}_{\mathbf{M},\Xi}(x-x^{\prime})$ can be seen as a product of the $\mathbf{M}$ and $\eta_0\Xi_{\lambda}^{\dag}\mathbf{M}^{\dag}\Xi_{\psi}^{\dag}\eta_0$ operators. In this program it becomes possible to define the amplitude of propagation for any other spinor.

This approach includes examples where the $\Xi$ operator is the same for particles and antiparticles. Nonetheless, if $\Xi_{particle}\neq\Xi_{antiparticle}$ then Eq.\eqref{propagadormapeado} can be split into a sum of two propagators, a term holding information related to the particle spinor and the remaining term holding information of the antiparticle spinor, given by\footnote{It is important to remark, if one take into account a more specific framework, like the adjoints defined in \cite{chengtipo42021,aaca}, thus, we may set the following relations
\begin{equation}\label{ultimaformadual}
\stackrel{\neg}{\lambda} = [\Xi_{\lambda}\lambda]^{\dag}\eta_0\mathcal{O}_{\lambda} \quad\mbox{and}\quad \stackrel{\neg}{\psi} = [\Xi_{\psi}\psi]^{\dag}\eta_0\mathcal{O}_{\psi}, 
\end{equation}
in which the $\mathcal{O}$ operator holds important physical information regarding the spin-sums. Thus, a more involved amplitude of propagation may emerge as
 \begin{eqnarray}
\mathcal{A}_{\mathbf{M},\Xi}(x-x^{\prime}) = [\mathbf{M}\mathcal{A}(x-x^{\prime})\mathcal{O}^{-1}_{\lambda}\eta_0\Xi_{\lambda}^{\dag}\mathbf{M}^{\dag}\Xi_{\psi}^{\dag}\eta_0\mathcal{O}_{\psi}]_{_{particle}}
\pm[\mathbf{M}\mathcal{A}(x-x^{\prime})\mathcal{O}^{-1}_{\lambda}\eta_0\Xi_{\lambda}^{\dag}\mathbf{M}^{\dag}\Xi_{\psi}^{\dag}\eta_0\mathcal{O}_{\psi}]_{_{anti-particle}}.
\end{eqnarray}
However, for the purposes of this work, we will not investigate this case further.}
 \begin{eqnarray}\label{propagadormapeado2}
\mathcal{A}_{\mathbf{M},\Xi}(x-x^{\prime}) = [\mathbf{M}\mathcal{A}(x-x^{\prime})\eta_0\Xi_{\lambda}^{\dag}\mathbf{M}^{\dag}\Xi_{\psi}^{\dag}\eta_0]_{_{particle}}\pm[\mathbf{M}\mathcal{A}(x-x^{\prime})\eta_0\Xi_{\lambda}^{\dag}\mathbf{M}^{\dag}\Xi_{\psi}^{\dag}\eta_0]_{_{anti-particle}}.
\end{eqnarray} 

Finally, according to \eqref{propagadormapeado2}, we easily obtain
\begin{eqnarray}\label{666}
\mathcal{A}(x-x^{\prime})= -i\xi \mathop{\mathrm{lim}}
\limits_{\epsilon\rightarrow 0^+}\Bigg\lbrace\Bigg[\mathbf{M}\int\frac{d^4 p}{(2\pi)^4}\frac{1}{2E(\boldsymbol{p})\upsilon}\Bigg(\frac{\sum_{h}\lambda_{h}(\boldsymbol{p})\stackrel{\neg}{\lambda}_{h}(\boldsymbol{p})(p_0+\sqrt{p_{j}p^{j}+m^2})}{p_{\mu}p^{\mu}-m^2+i\epsilon}
\Bigg)e^{-ip_{\mu}(x^{\mu}-x^{\prime\mu})}\boldsymbol{\Delta}\Bigg]_{_{particle}}\nonumber\\
\pm \Bigg[\mathbf{M}\int\frac{d^4 p}{(2\pi)^4}\frac{1}{2E(\boldsymbol{p})\upsilon}\Bigg(\frac{\sum_{h}\chi_{h}(-\boldsymbol{p})\stackrel{\neg}{\chi}_{h}(-\boldsymbol{p})(p_0-\sqrt{p_{j}p^{j}+m^2})}{p_{\mu}p^{\mu}-m^2+i\epsilon} \Bigg)e^{-ip_{\mu}(x^{\mu}-x^{\prime\mu})}\boldsymbol{\Delta}\Bigg]_{_{anti-particle}}\Bigg\rbrace, 
\end{eqnarray}
in which we have defined $\boldsymbol{\Delta}=\eta_0\Xi_{\lambda}^{\dag}\mathbf{M}^{\dag}\Xi_{\psi}^{\dag}\eta_0$.

We finally derived the most general form of the amplitude of propagation. For fermionic particles we have that we can write $\boldsymbol{\Delta}_{F}=\gamma_0\Xi_{\lambda}^{\dag}\mathbf{M}^{\dag}\Xi_{\psi}^{\dag}\gamma_0$ while for spin-$1/2$ bosonic particles $\boldsymbol{\Delta}_{B}=a_0\Xi_{\lambda}^{\dag}\mathbf{M}^{\dag}\Xi_{\psi}^{\dag}a_0$.

Notice that the general amplitude of propagation $\mathcal{A}_{\mathbf{M},\Xi}(x-x^{\prime})$ is easily obtained from a multiplication of matrix operators. Taking into account the relations previously introduced in \eqref{gamma-a}, we are able to define $\boldsymbol{\Delta}=\gamma_5\gamma_0\Xi_{\lambda}^{\dag}\mathbf{M}^{\dag}\Xi_{\psi}^{\dag}\gamma_0\gamma_5$, remembering the constant nature of $\mathbf{M}$ (furnishing $\gamma_5\mathbf{M}\gamma_5 = \mathbf{M}$). In other words $\boldsymbol{\Delta}_{B} = \gamma_5\boldsymbol{\Delta}_{F}\gamma_5$ and therefore
\begin{eqnarray*}
\mathcal{A}(x-x^{\prime})= -i\xi \mathop{\mathrm{lim}}
\limits_{\epsilon\rightarrow 0^+}\Bigg\lbrace\gamma_5\Bigg[\mathbf{M}\int\frac{d^4 p}{(2\pi)^4}\frac{1}{2E(\boldsymbol{p})\upsilon}\Bigg(\frac{\sum_{h}\lambda_{h}(\boldsymbol{p})\stackrel{\neg}{\lambda}_{h}(\boldsymbol{p})(p_0+\sqrt{p_{j}p^{j}+m^2})}{p_{\mu}p^{\mu}-m^2+i\epsilon}
\Bigg)e^{-ip_{\mu}(x^{\mu}-x^{\prime\mu})}\Delta_{F}\Bigg]_{_{_{particle}}}\!\!\!\!\!\!\!\!\!\!\!\!\!\!\!\!\gamma_5\nonumber\\
\pm  \gamma_5\Bigg[\mathbf{M}\int\frac{d^4 p}{(2\pi)^4}\frac{1}{2E(\boldsymbol{p})\upsilon}\Bigg(\frac{\sum_{h}\chi_{h}(-\boldsymbol{p})\stackrel{\neg}{\chi}_{h}(-\boldsymbol{p})(p_0-\sqrt{p_{j}p^{j}+m^2})}{p_{\mu}p^{\mu}-m^2+i\epsilon} \Bigg)e^{-ip_{\mu}(x^{\mu}-x^{\prime\mu})}\Delta_{F}\Bigg]_{_{anti-particle}}\gamma_5 \Bigg\rbrace, 
\end{eqnarray*}
or in a more compact form
\begin{equation}
\mathcal{A}_{B}(x-x^{\prime}) = \gamma_5\mathbf{M}\mathcal{A}_{F}(x-x^{\prime})\boldsymbol{\Delta}_{F} \gamma_5,
\end{equation}
allowing to explicitly show the connection among bosonic and fermionic amplitudes of propagation via the chiral operator $\gamma_5$. This approach allows to connect different classes of spin-half particles hence determining the structure of the propagator of some spinor classes once we know the structure of the propagator of other spinor classes \cite{rodolfospinhalf,rodolfobosonsphases}.
\subsection{Fields ruled by a first order equation: On the spin-half bosons framework}\label{sect3A}
Now, for the sake of demonstration, in this section we will show the consistency of the results previously introduced in section \ref{generalpropagatorboson} with a case already established in \cite{dharamboson}. To do so, we will then map $\lambda $ to a spinor, which belongs to class II of the classification presented in \cite{rodolfobosonsphases}. So we define for particles $\mathbf{M} =\mathrm{diag}(1, 1, 1, 1)$ and for antiparticles: $\mathbf{M} = \mathrm{diag}(-1, -1, 1, 1)$, $\Xi_{\lambda} = \Xi_{\psi} = \mathbbm{I}$ leading to $\boldsymbol{\Delta} = \mathrm{diag}(1, 1, 1, 1)$ for particles and $\boldsymbol{\Delta} = \mathrm{diag}(1, 1, -1, -1)$ for anti-particles. Bearing in mind all the observations above, we proceed with the amplitude of propagation computation 
\begin{eqnarray}\label{ampli1}
\mathcal{A}(x-x^{\prime})= -i\xi\Bigg\lbrace \Bigg[\mathbf{M}\int\frac{d^4 p}{(2\pi)^4}\frac{1}{2E(\boldsymbol{p})\upsilon}\Bigg(\frac{(a_{\mu}p^{\mu}+m\mathbbm{I})(p_0+\sqrt{p_{j}p^{j}+m^2})}{p_{\mu}p^{\mu}-m^2+i\epsilon}
\Bigg)e^{-ip_{\mu}(x^{\mu}-x^{\prime\mu})}\boldsymbol{\Delta}\Bigg]_{_{particle}}\nonumber\\
\pm  \Bigg[\mathbf{M}\int\frac{d^4 p}{(2\pi)^4}\frac{1}{2E(\boldsymbol{p})\upsilon}\Bigg(\frac{(-a_{\mu}p^{\mu}+m\mathbbm{I})(p_0-\sqrt{p_{j}p^{j}+m^2})}{p_{\mu}p^{\mu}-m^2+i\epsilon} \Bigg)e^{-ip_{\mu}(x^{\mu}-x^{\prime\mu})}\boldsymbol{\Delta}\Bigg]_{_{anti-particle}}\Bigg\rbrace, 
\end{eqnarray}
the terms in first integral in \eqref{ampli1} providing
\begin{equation}
[\mathbf{M}(a_{\mu}p^{\mu}+m\mathbbm{I})\boldsymbol{\Delta}]_{_{anti-particles}} = (a_{0}p^{0}+\boldsymbol{a}\cdot\boldsymbol{p}+m\mathbbm{I}),
\end{equation}
with the second term giving
\begin{equation}
[\mathbf{M}(-a_{\mu}p^{\mu}+m\mathbbm{I})\mid_{_{\boldsymbol{p}\rightarrow - \boldsymbol{p}}}\boldsymbol{\Delta}]_{_{anti-particles}} = (-a_{0}p^{0}-\boldsymbol{a}\cdot\boldsymbol{p}-m\mathbbm{I}),
\end{equation}
and leading, eventually, to
\begin{eqnarray}
\mathcal{A}(x-x^{\prime})= -i\xi\Bigg\lbrace \Bigg[\int\frac{d^4 p}{(2\pi)^4}\frac{1}{2E(\boldsymbol{p})\upsilon}\Bigg(\frac{(a_{0}p^{0}+\boldsymbol{a}\cdot\boldsymbol{p}+m\mathbbm{I})(p_0+\sqrt{p_{j}p^{j}+m^2})}{p_{\mu}p^{\mu}-m^2+i\epsilon}
\Bigg)e^{-ip_{\mu}(x^{\mu}-x^{\prime\mu})}\Bigg]\nonumber\\
\pm \Bigg[\int\frac{d^4 p}{(2\pi)^4}\frac{1}{2E(\boldsymbol{p})\upsilon}\Bigg(\frac{(-a_{0}p^{0}-\boldsymbol{a}\cdot\boldsymbol{p}-m\mathbbm{I})(p_0-\sqrt{p_{j}p^{j}+m^2})}{p_{\mu}p^{\mu}-m^2+i\epsilon} \Bigg)e^{-ip_{\mu}(x^{\mu}-x^{\prime\mu})}\Bigg]\Bigg\rbrace, 
\end{eqnarray}

For consistency (positive orthonormal relation and also due the structure introduced in \eqref{ssbosonanti}) we are forced to pick the plus sign. This is equivalent to the choice \eqref{comutador}. After some algebra, we reach the following
\begin{eqnarray}
\mathcal{A}(x-x^{\prime})= -i\xi \int\frac{d^4 p}{(2\pi)^4}\frac{1}{\upsilon}\frac{a_{\mu}p^{\mu}+m\mathbbm{I}}{p_{\mu}p^{\mu}-m^2+i\epsilon}
e^{-ip_{\mu}(x^{\mu}-x^{\prime\mu})}.
\end{eqnarray}

Accordingly \cite{dharamboson}, the choice $\xi=i$ and $\upsilon=1$ makes $\mathcal{A}(x\to x^\prime)$ to be the Green's function of the differential operator 
\begin{eqnarray}
( i a_{\mu^\prime}\partial^{\mu^\prime}-m\mathbbm{I} )
\mathcal{A}(x\to x^\prime)   &=&  i \xi\delta^4(x^\prime-x),
\nonumber\\
&=&-\delta^4(x^\prime-x).
\end{eqnarray}
Then, we can define the Feynman-Dyson propagator for the \emph{fields which obey a first-order derivative equation} to be
\begin{equation}\label{propagatorbosonfinal}
\mathcal{S}_{\textrm{FD}}(x-x^\prime)=  \int\frac{\text{d}^4 p}{(2 \pi)^4}\,
e^{-i p_\mu(x^{\prime\mu}-x^\mu)}
\frac{a_\mu p^\mu + m\mathbbm{I}}{p_\mu p^\mu -m^2 + i\epsilon}. 
\end{equation}
Analogously, the Dirac counterpart reads
\begin{equation}\label{propagatordiracfinal}
\mathcal{S}_{\textrm{FD}}(x-x^\prime)=  \int\frac{\text{d}^4 p}{(2 \pi)^4}\,
e^{-i p_\mu(x^{\prime\mu}-x^\mu)}
\frac{\gamma_{\mu}p^{\mu} + m\mathbbm{I}}{p_\mu p^\mu -m^2 + i\epsilon},  
\end{equation}
as expected for eigenspinors of $a_{\mu}p^{\mu}$ and $\gamma_{\mu}p^{\mu}$ operators.

\subsection{Fields ruled by Klein-Gordon equation: On the mass-dimension-one fermions framework}
Guided by the existing examples on mass-dimension $1$ fermions, now we are able to define the most general amplitude of propagation and the propagator itself.

Considering the existing examples in \cite{mdobook,dharamnewfermions} we can define the following: $\Xi_{Elko}=m^{-1}\mathcal{G}(\phi)\gamma_{\mu}p^{\mu}$ \cite{mdobook} as well as $\Xi_{Ah-7}=m^{-1}\gamma_{\mu}p^{\mu}$ for particles and $\Xi_{Ah-7}=-m^{-1}\gamma_{\mu}p^{\mu}$ for anti-particles\footnote{Where the sub-index Ah-7 means Ahluwalia class-7 spinors.} \cite{dharamnewfermions}. We also have for particles $\mathbf{M}= \mathrm{diag}(1, 1, 0, 0)$ and consequently for antiparticles $\mathbf{M}= \mathrm{diag}(0, 0, 1, 1)$. Moreover, taking into account \eqref{666}, we are able to write the following
\begin{equation}
[\mathbf{M}2m\mathbbm{I}\boldsymbol{\Delta}]_{_{particles}} = 2m\mathbbm{I},
\end{equation}
and 
\begin{equation}
[\mathbf{M}(-2m)\mathbbm{I}\boldsymbol{\Delta}]_{_{anti-particles}} = 2m\mathbbm{I},
\end{equation}
where $\mathbbm{I}$ stands for the $4\times 4$ identity matrix. Therefore
\begin{eqnarray}
\mathcal{A}(x-x^{\prime})&=& -i\xi \Bigg\lbrace\Bigg[\int\frac{d^4 p}{(2\pi)^4}\frac{1}{2E(\boldsymbol{p})\upsilon}\Bigg(\frac{2m\tilde{\mathbbm{I}}(p_0+\sqrt{p_{j}p^{j}+m^2})}{p_{\mu}p^{\mu}-m^2+i\epsilon}
\Bigg)e^{-ip_{\mu}(x^{\mu}-x^{\prime\mu})}\Bigg]\nonumber\\
&\pm & \Bigg[\int\frac{d^4 p}{(2\pi)^4}\frac{1}{2E(\boldsymbol{p})\upsilon}\Bigg(\frac{2m\mathbbm{I}(p_0-\sqrt{p_{j}p^{j}+m^2})}{p_{\mu}p^{\mu}-m^2+i\epsilon} \Bigg)e^{-ip_{\mu}(x^{\mu}-x^{\prime\mu})}\Bigg]\Bigg\rbrace.
\end{eqnarray}

Again, consistency forces the choice of the minus sign in \eqref{general-propagator-form}. This is equivalent to \eqref{anticomutador}. Setting $\upsilon = m$ gives
\begin{equation}
\mathcal{A}(x\to x^\prime)  = -i 2 \xi \int\frac{\text{d}^4 p}{(2 \pi)^4}\,
e^{-i p_\mu(x^{\prime\mu}-x^\mu)}
\frac{\mathbbm{I}}{p_\mu p^\mu -m^2 + i\epsilon}.\label{eq:AmplitudeWithXi}
\end{equation}

Then, the normalisation of the $\xi$ factor leads to
\begin{equation}
\xi = \frac{i m^2}{2},
\end{equation}
yielding
\begin{equation}
\mathcal{A}(x\to x^\prime)  =  m^2 \int\frac{\text{d}^4 p}{(2 \pi)^4}\,
e^{-i p_\mu(x^{\prime\mu}-x^\mu)}
\frac{\mathbbm{I}}{p_\mu p^\mu -m^2 + i\epsilon},
\end{equation}
so that
\begin{equation}
\left(\partial_{\mu^\prime} \partial^{\mu^\prime} \mathbbm{I} + m^2\mathbbm{I}\right)
S_{\textrm{FD}}(x^\prime-x)  = -  \delta^4(x^\prime - x).
\end{equation}
The Feynman-Dyson propagator for fields obeying a second-order equation is 
\begin{align}\label{propagatormdofinal}
S_{\textrm{FD}}(x^\prime-x) & \stackrel{\textrm{def}}{=}  \frac{1}{m^2} 
\mathcal{A}(x\to x^\prime)\nonumber\\
 &= \int\frac{\text{d}^4 p}{(2 \pi)^4}\,
e^{-i p_\mu(x^{\prime\mu}-x^\mu)}
\frac{\mathbbm{I}}{p_\mu p^\mu -m^2 + i\epsilon}.
\end{align}
as expected for mass-dimension-one fermions. The results above hold true for all spinors carrying a more involved dual structure, as presented in \eqref{map2}, \eqref{map2.2} and \eqref{ultimaformadual}.

\section{Concluding Remarks and Outlooks}\label{concluding}
In the present work, we have shown how to construct the most general form of the propagator for fermionic and bosonic fields carrying generalized dual adjoints and governed by non-standard statistics.

Such a result is general because it is defined upon a mapping procedure among spinors and dual structures. This mechanism brings to light the possibility to write the core of the propagator, namely the spin-sum relations, for any (class of) spinor. Thus, once we have the general form of the propagator, we may find the connection (mapping operator) among different classes of spinors and, then, finally write the propagator for a specific spinor theory.  

As clear, the propagator is closely connected to the spinor dynamics. Generally, mass-dimension $1$ fermions have a similar structure to the scalar field propagator, whereas spin-$1/2$ bosons (as well as Dirac fermions) are governed by a first-order dynamical equation (which stands for eigenspinors of the operator $a_{\mu}p^{\mu}$ and $\gamma_{\mu}p^{\mu}$).

Finally, we showed the possibility to connect the Dirac and spin-$1/2$ bosons propagators in terms of a relationship between the bosonic and fermionic representations of Clifford algebra.

\appendix

\section{Some results based on new dual definitions}

Quite recently, some new definitions of dual structure opened windows to a new interpretation of how spin-half bosons emerge (from Dirac spinors) and also how they evade the spin-statistic theorem \cite{dharam2022spin}. As claimed, the new dual structure furnishes a local and Lorentz-invariant theory and also provide a positive-definite Hamiltonian. Such aforementioned features are carried by a new dual structure, which reads
\begin{eqnarray}
\stackrel{\neg}{\lambda} =  +\bar{\lambda}^{\dag}\gamma_0, \quad \mbox{and}
\stackrel{\neg}{\chi} =  -s\bar{\chi}^{\dag}\gamma_0,\label{dual2022}
\end{eqnarray}
in which $s=1$ stands for fermionic field and $s=-1$ stands for bosonic field. Bearing in mind the dual definition in \eqref{dual2022}, the spin sums \eqref{spinsums1} and \eqref{spinsums2} are now replaced by the following set 
  \begin{eqnarray}
\sum_{particle}\lambda_{i}(\p)\gdualn\lambda_{i}(\p) &=& \gamma_{\mu}p^{\mu}+ m\mathbbm{I}, 
\\
\sum_{anti-particle}\!\!\!\!\!\!\chi_{i}(\p)\gdualn\chi_{i}(\p) &=& -(\gamma_{\mu}p^{\mu}- m\mathbbm{I}).
\end{eqnarray}
With these new results at hands, Eq.\eqref{666} is written, accordingly \cite[page3]{dharam2022spin}, as
\begin{eqnarray}\label{666new}
\mathcal{A}(x-x^{\prime})= -i\xi \mathop{\mathrm{lim}}
\limits_{\epsilon\rightarrow 0^+}\Bigg\lbrace\Bigg[\mathbf{M}\int\frac{d^4 p}{(2\pi)^4}\frac{1}{2E(\boldsymbol{p})\upsilon}\Bigg(\frac{\sum_{h}\lambda_{h}(\boldsymbol{p})\stackrel{\neg}{\lambda}_{h}(\boldsymbol{p})(p_0+\sqrt{p_{j}p^{j}+m^2})}{p_{\mu}p^{\mu}-m^2+i\epsilon}
\Bigg)e^{-ip_{\mu}(x^{\mu}-x^{\prime\mu})}\boldsymbol{\Delta}\Bigg]_{_{particle}}\nonumber\\
+ \Bigg[\mathbf{M}\int\frac{d^4 p}{(2\pi)^4}\frac{1}{2E(\boldsymbol{p})\upsilon}\Bigg(s\frac{\sum_{h}\chi_{h}(-\boldsymbol{p})\stackrel{\neg}{\chi}_{h}(-\boldsymbol{p})(p_0-\sqrt{p_{j}p^{j}+m^2})}{p_{\mu}p^{\mu}-m^2+i\epsilon} \Bigg)e^{-ip_{\mu}(x^{\mu}-x^{\prime\mu})}\boldsymbol{\Delta}\Bigg]_{_{anti-particle}}\Bigg\rbrace, 
\end{eqnarray}
According to the type of field, the statistic will be dictated by the $s$ parameter. If the field is to be fermionic, $s=1$, forcing the anticommutative relations among the creator and annihilator operators. If the field is bosonic, $s=-1$, and commutative relations must be taken into account. For more details, we refer to \cite{dharam2022spin}.

Now, following the very same steps as in Sect.\eqref{sect3A}, and bearing in mind the correct relation among the creator/annihilator operators as well as the value of $s$, the above amplitude of propagation yields
\begin{equation}
\mathcal{S}_{\textrm{FD}}(x-x^\prime)=  \int\frac{\text{d}^4 p}{(2 \pi)^4}\,
e^{-i p_\mu(x^{\prime\mu}-x^\mu)}
\frac{\gamma_\mu p^\mu + m\mathbbm{I}}{p_\mu p^\mu -m^2 + i\epsilon}. 
\end{equation}
as expected for Dirac fermions and also for spin-half Therefore, we show the similarity of both results presented in \cite{dharamboson} and \cite{dharam2022spin}.
\bibliographystyle{unsrt}
\bibliography{refs}
\end{document}